# The dermal-epidermal junction instructs epidermal keratinocytes

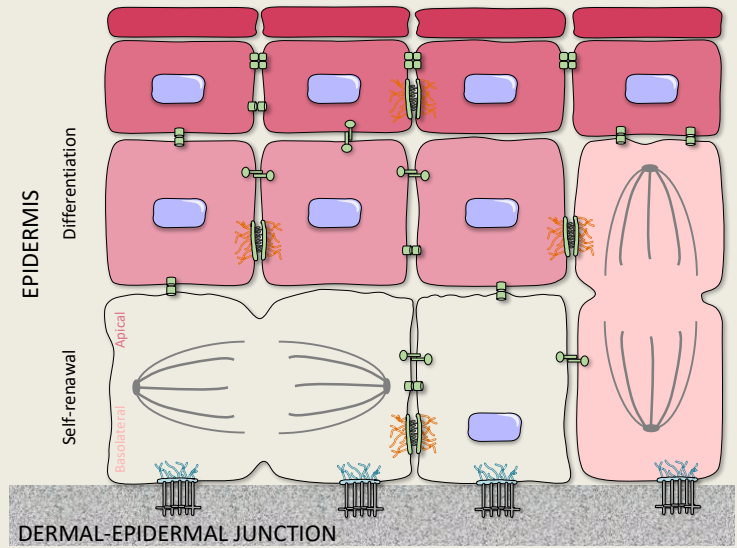

## CONCLUSION

The function of the epidermis relies on mechanical, structural, and biochemical support from the underlying specialized basement membrane named dermal epidermal junction.

# The basement membrane in epidermal polarity, stemness and regeneration.

Patricia Rousselle*, Chloé Laigle, Gaelle Rousselet.

Laboratoire de Biologie Tissulaire et Ingénierie Thérapeutique, UMR 5305; CNRS; Univ. Lyon 1; 7 passage du Vercors, 69367, Lyon, France

*Correspondance : Patricia Rousselle, LBTI, 7 passage du Vercors – 69367 Lyon Cedex 07, France. Tel: (33) 4 72 72 26 39; Fax: (33) 4 72 72 26 02; E-mail: patricia.rousselle@cnrs.fr

Running title: Basement membrane in epidermal homeostasis and repair

Abbreviations: atypical protein kinase C (aPKC), basement membrane (BM), dermal-epidermal junction (DEJ), epidermal growth factor receptor (EGF-R); epidermolysis bullosa (EB), extracellular signal-regulated kinase (ERK), extracellular matrix (ECM), heparan sulphate (HS), integrin-linked kinase (ILK), laminin (LM), melanoma chondroitin sulphate proteoglycan (MCSP), mitogen activated protein kinase (MAPK), partitioning defective (Par), phosphatidyl inositol-3 kinase (PI-3K), phospholipid phosphatidylinositol 4,5-bisphosphate (PIP2), three-dimensional (3D).

Abstract

The epidermis is a specialized epithelium that constitutes the outermost layer of the skin, and it provides a protective barrier against environmental assaults. Primarily consisting of multi-layered keratinocytes, the epidermis is continuously renewed by proliferation of stem cells and the differentiation of their progeny, which undergo terminal differentiation as they leave the basal layer and move upward toward the surface, where they die and slough off. Basal keratinocytes rest on a basement membrane at the dermal-epidermal junction that is composed of specific extracellular matrix proteins organized into interactive and mechanically supportive networks. Firm attachment of basal keratinocytes, and their dynamic regulation via focal adhesions and hemidesmosomes, are essential for maintaining major skin processes, such as self-renewal, barrier function, and resistance to physical and chemical stresses. The adhesive integrin receptors expressed by epidermal cells serve structural, signaling, and mechanosensory roles that are critical for epidermal cell anchorage and tissue homeostasis. More specifically, the basement membrane components play key roles in preserving the stem cell pool, and establishing cell polarity cues enabling asymmetric cell divisions, which result in the transition from a proliferative basal cell layer to suprabasal cells committed to terminal differentiation. Finally, through a well-regulated sequence of synthesis and remodeling, the components of the dermal-epidermal junction play an essential role in regeneration of the epidermis during skin healing. Here too, they provide biological and mechanical signals that are essential to the restoration of barrier function.

**Keywords:** epidermis; basement membrane; dermal-epidermal junction, extracellular matrix; polarity; stem cell; regeneration; aging.

*Introduction*

Skin is the protective barrier that shields the body from environmental insults (1). The outermost layer of the skin, the epidermis (epithelium), is separated from the dermis (mesenchyme) by a thin layer of highly organized extracellular matrix (ECM) proteins that comprise the dermal-epidermal junction (DEJ). This type of ECM is categorized as a supramolecular matrix arrangement termed basement membrane (BM), to which are added anchoring proteins that serve to reinforce dermo-epidermal cohesion. BMs compartmentalize, provide structural support, and regulate cell behavior through signaling activities that influence tissue growth, guide cell migration, and support cell survival (2, 3). Largely populated by keratinocytes, the epidermis undergoes constant turnover as epidermal stem cells of the basal layer, which have a high proliferative potential, constantly generate new daughter cells or transit-amplifying cells. After a few cycles of division, daughter cells begin a process of terminal differentiation (cornification), leading to the formation of the stratum corneum (4). Epidermal integrity and renewal depend on the balance between proliferation and differentiation of the basal layer of keratinocytes that are in contact with the BM. Additionally, various growth factors, morphogens, and other regulatory macromolecules tethered to the BM provide signals involved in governing keratinocyte adhesion, differentiation, stratification, and survival (5, 6).

One particularity of the DEJ is that it includes very sophisticated anchoring proteins, which reinforce the cohesion between dermis and epidermis, and thereby increase the epidermis' resistance to the mechanical forces to which the skin is constantly subjected. These anchoring complexes link basal keratinocytes via attachment dots called hemidesmosomes that span the BM and anchor the dermis through specific elements called anchoring fibrils. These essential structures contain specific components, such as α6β4 integrin, laminin (LM) isoform 332, and collagens VII and XVII (7). The DEJ is formed through contributions from both epidermal and dermal cells (8), and it exhibits a distinctive microarchitecture characterized by an undulating pattern due to epidermal projections into the dermis that intercalate between dermal papillae (7, 9). This BM geometry increases the surface area of the DEJ, strengthens the dermal–epidermal connectivity, and imposes mechanical constraints on keratinocytes with consequences on their fate (9). Thus, the function of the epidermis relies on mechanical, structural, and biochemical support from the underlying DEJ.

**STRUCTURAL ORGANIZATION OF THE DERMAL-EPIDERMAL JUNCTION**

Within the skin, the DEJ provides essential structural support to the adjacent basal epidermal keratinocytes (10, 11), constituting a specific informative niche that mediates signals and plays a preponderant role in epidermal renewal (12). The basic structure of the DEJ can be viewed as a BM assembled according to the commonly accepted model (5, 8, 13). The BM framework is comprised of two independent networks of polymerized LMs and cross-linked collagen IV that are interconnected and stabilized by the glycoprotein nidogen and the heparan sulphate (HS) proteoglycan, perlecan (Fig. 1). Collagen IV is the most abundant structural component of the BM, and it is largely responsible for its mechanical properties (14, 15). The combination of three α collagen chains, α1α1α2(IV) is the most abundant collagen IV molecule in all BMs (16), while the α5α5α6(IV) isoform is also present in the DEJ (17). Collagen IV α chains contain an N-terminal 7S domain, a central collagen domain, and a C-terminal non-collagenous (NC)1 domain (18). This structure enables collagen IV molecules to self-aggregate into dimers through their NC1 domains, and into tetramers through their N-terminal 7S domains, leading to formation of a hexameric network (Fig. 1) (19). It is believed that the covalent crosslinks of the collagen IV meshed network are central to the mechanical properties of the BM (20-22).

Although collagen IV may be essential for BM integrity, it is not critical to its initial formation (23). Indeed, LM is considered the initiator of BM assembly at the basolateral surface of cells. Secreted LMs are comprised of α, β, and γ chains assembled into a heterotrimer that has an asymmetrical, cruciform shape (Fig. 1) (24). cDNA sequencing has led to identification of the primary structures of five LM α chains (α1, α2, α3, α4, and α5), four β chains (β1, β2, β3, and β4), and three γ chains (γ1, γ2, and γ3), yielding at least 18 heterotrimeric isoforms (25). Moreover, due to alternative splicing, the gene encoding the α3 chain produces two splice variants: a short-α3A chain and a long-α3B. As long as each of the three constituent LM chains includes an N-terminal LN domain, LMs can self-associate into polymeric, sheet-like networks (26-28). These LM networks are mainly non-covalent in nature, allowing dynamic interactions. To initiate BM assembly, the laminin G-type (LG) modules LG1–LG5 at the C-terminus of the α chain, which protrude from the LM long arm, bind directly to cells through cell surface receptors: integrins, α-dystroglycan, and sulfated glycolipids (29, 30). This property enables recruitment of other BM components, including nidogen, collagen IV, and perlecan, in a temporal hierarchy (Fig. 1) (31, 32). The LM network of the epidermal BM results from self-assembly of the heterotrimer α5β1γ1 or LM-511. Various analyses have revealed that the supramolecular organization of BMs is asymmetrical and bifunctional. The two faces have different exposed proteins: a LM-rich epithelial side and a collagen-rich stromal side (33). Atomic force microscopic studies have revealed that BM polarization is characterized by distinct mechanical properties, the epithelial side being 2-4 times stiffer than the stromal side (33).

Two other LM isoforms found in the DEJ, LM-332 (α3Aβ3γ2) and LM-311 (α3Aβ1γ1) (34-36), are unable to self-assemble because their constitutive α3 and/or γ2 chains are truncated at the N-terminal end, such that they lack the required LN domains. However, networks containing LM-332 have been purified from skin, indicating that some form of LM assembly occurs (37). Their integration follows a different type of assembly pattern specific to BMs underlying epithelia that are subjected to strong mechanical forces, such as the DEJ (36, 38). For integration into the DEJ, LM-332 first cross-links with LM-311, forming a complex better equipped for integration (34, 39). In addition, both monomeric LM-332 and the LM-332/311 dimer associate with the N-terminal NC1 domain of collagen VII (40, 41). Collagen VII homotrimers α1α1α1(VII) dimerize and form fibrillar structures, called anchoring fibrils, which firmly connect the DEJ to the superficial dermis. Dermal epidermal cohesion depends on the interaction that occurs within the LN domain of the LM β3 chain and the fibronectin-like III repeats within the NC1 domain of collagen VII (42, 43). LM-332 is critical for dermo-epidermal anchorage, because it associates with basal epidermal keratinocytes through its C-terminal domains, initiating and maintaining hemidesmosomes, and binding to collagen VII through its N-terminus (7, 36). Hemidesmosomes are highly organized adhesion structures formed in the basal layer of the epidermis, where they link the BM with the keratin-containing intermediate filament cytoskeleton, providing mechanical strength and durability (44). Both LM-332/311 and LM-511 are secreted from keratinocytes, and they incorporate into the BM zone of hair follicles and interfollicular epidermis. Another component of this specialized cell–BM interface category is collagen XVII, a transmembrane collagen homotrimer α1α1α1(XVII) (45). Collagen XVII plays a crucial role in keratinocyte anchorage by binding extracellular LM-332, and by controlling intracellular molecular assemblies critical for hemidesmosome assembly and stability (46, 47). Finally, collagen XVIII, which belongs to the multiplexin (multiple-helix domains with interruptions) collagen subgroup, is believed to reinforce DEJ stability and mechanical properties through its numerous interactive and bridging properties (48).

**THE DERMAL-EPIDERMAL JUNCTION PROVIDES ADHESION AND SIGNALING CUES SUPPORTING STEMNESS**

Integrin-mediated cell-matrix adhesions play a key role in linking the cytoskeleton to the BM. Within the epidermis, the most abundant, constitutive transmembrane integrins are α2β1, α3β1, and α6β4 (49). The α2β1 integrin is observed at cell–cell junctions (50), where it is proposed to be a ligand for collagen XXIII in the basal layer (51). More recently, its clustering at keratinocyte cell–cell adhesions was demonstrated to coordinate the stabilization of adherens junctions (52).

Basal keratinocytes closely interact with the BM through both integrin α3β1-enriched focal adhesions and α6β4-induced hemidesmosomes. Intracellularly, these two integrins interact with and respectively regulate the actin and keratin cytoskeletons via a multitude of scaffold and regulatory proteins (53, 54). Firm keratinocyte attachment and dynamic regulation via focal adhesions and hemidesmosomes are essential to maintain major skin functions, such as self-renewal, barrier function, and resistance to physical and chemical stresses. The C-terminal G domain of the α3 chain in LM-332 is the ligand for both the α3β1 and α6β4 integrins in keratinocytes, and it can promote the formation of these two different types of attachment structures (55-57). Focal adhesions allow dynamic, through reversible associations with integrins, while hemidesmosomes form stable, mechanically-resistant adhesion complexes. Upon binding to LM-332, hemidesmosomes assemble through the interaction of the β4 integrin cytoplasmic tail with collagen XVII and plectin, which binds to intermediate filaments in the cytoplasm (58). Recent data highlight the close connections between these two types of adhesion structures, particularly during keratinocyte migration, when hemidesmosomes go through dynamic remodeling (59, 60).

Hemidesmosomes are also considered signaling platforms. Following activation of the epidermal growth factor receptor (EGF-R), the β4 integrin subunit cytoplasmic tail is phosphorylated, leading to the extracellular signal-regulated kinase (ERK)/mitogen-activated protein kinase (MAPK) and phosphatidyl inositol-3 kinase (PI-3K) signaling cascade engagement that supports growth and survival (54, 61, 62). In this manner, junctional and cytoskeletal proteins control the maintenance of the stem cell population that resides in the epidermal basal layer. Hemidesmosome integrity, associated with sufficient collagen XVII expression in basal keratinocytes, guarantees the potential for symmetrical stem cell division and epidermal homeostasis (63). It was recently proposed that collagen XVII connection with the atypical protein kinase C (aPKC)-partitioning defective (Par) complex may be a mechanism for regulating epidermal cell polarity (64).

LM-332 is predominantly expressed in the BM underlying $CD271^+$ human interfollicular stem cells at the top of the dermal papilla, where this protein has been suggested to impact keratinocyte differentiation (65). Conditional deletion of LM-332 in the epidermis of adult mice profoundly disturbs the keratinocyte differentiation program, leading to exaggerated expression of terminal differentiation markers and major F-actin-related cell shape anomalies—a phenotype indicating disrupted mechanosensitive signaling pathways (66). Furthermore, data from junctional epidermolysis bullosa patients indicate that epidermal stem cell preservation requires a LM-332/α6β4 integrin-induced adhesion-dependent mechanical pathway to support the activity of the mechanosensitive transcriptional regulator Yes-associated protein (YAP) (67). Interestingly, *Caenorhabditis elegans* hemidesmosomes, which are structurally and functionally analogous to vertebrate hemidesmosomes, have also been identified as mechanosensors (68). In humans, LM-332/integrin β4/plectin association reportedly plays a role in the regulation of cellular mechanical forces through mechanosensitive cross-talk with focal adhesions (60). These elements demonstrate that, in addition to being a mainstay in epidermal anchorage, LM-332 signaling also strongly impacts epidermal homeostasis. Notably, LM-332 research goes far beyond the context of skin biology, and its effect on cancer stem cell quiescence has been addressed in other recent reviews (69, 70).

Accumulating data indicate that LM-511 is involved in the status and fate of epidermal stem cells. Expressed along the epidermal BM, LM-511 is particularly abundant in the BM zone surrounding hair follicles. Studies in mice have reported that LM-511 is essential for hair follicle elongation and downward growth into the dermis through its interaction with α3β1 integrin (71-76). Maintenance of the LM-511/LM-332 expression proportions and patterns in the hair follicle BM is crucial for regulating the epidermal stem cell niche and hair growth (77). It has also been proposed that LM-511 plays a role in the maintenance of interfollicular epidermal stem/progenitor cells (78), and keratinocytes expression of LM-511 correlates with increased capacity for epidermal regeneration (79). LM-511 has been used as a matrix for feeder-free growth of pluripotent cells (80) and epithelial cells (81), and the physiological presentation of cell-secreted proteins interestingly appears to be more efficient, suggesting that the organisation of LM-511 after deposition within the matrix is essential (82, 83). Previous data showed that direct binding of mouse embryonic stem cells to LM-511 can maintain these cells in a quiescent state (84). The function of LM-511 in the interfollicular epidermis is not clear. An early report of the characterisation of LM-α5 subunit expression in mice reported its association with epithelial cell maturation, implicating a role for this LM in the maintenance of differentiated epithelia (85). Integrins α5β1, α6β1 and αvβ1/β3 and the non-integrin receptor Lutheran blood group antigen have been shown to bind LM-511, however their relevance to keratinocyte biology is not known (86-89). The development of a LM-α5 chain keratinocyte-specific knockout mouse provided some interesting answers to this question (90). In addition to an expected delayed/absent hair outgrowth, mice showed no changes in LM-332 expression and hemidesmosome organization. Loss of LM-α5 in the epidermal BM was associated with epidermal hyperproliferation, increased expression of dermally derived ECM molecules, and elevated immune cell infiltration into the dermis. In-depth examination of this phenotype led to the hypothesis that LM-511 not only prevents keratinocyte proliferation and migration but may also control the morphogenic environment around the BM (90).

Together, these data highlight that LMs and LM-binding integrins found in the epidermis are structural, signaling, and mechanosensory receptors that are critical for epidermal cell adhesion and homeostasis.

## IMPACT OF THE BASEMENT MEMBRANE ON EPITHELIAL POLARITY

Interactions between basal keratinocytes and the BM provide epithelia with survival, proliferation, and differentiation signals, as well as directional cues to establish polarity. The polarity of epithelium results from apical and basal differential compartmentalization, intercellular adhesions, and mitotic spindle orientation to spatially direct asymmetric division (91, 92). In general, polarized epithelial cells form a continuous layer in which cells are connected to each other by tight and adherens junctions, creating a barrier to the outward environment (Fig. 2A). In many epithelia, the apical surface of the cell faces the external environment, while the basal surface is attached to the BM. Tight junctions comprised of peripheral and transmembrane proteins form a strong seal between adjacent cells at the apical surface, thereby controlling the selective diffusion of solutes, ions, and proteins (93). They also form a boundary between the apical and basal surfaces of cells, thus preventing the vertical diffusion of membrane proteins between poles. Adherens junctions classically comprise transmembrane cadherins and peripheral catenins (94). These junctions form homophilic intercellular interactions and intracellularly connect to the actin cytoskeleton.

Polarity establishment and maintenance in epithelial cells is influenced by activation of the small GTPases Rac1 and Cdc42, as well as by three evolutionary conserved groups of cytoplasmic polarization signaling proteins: the Par system, the Crumbs/Pals1/PATJ complex, and the Scribble (Scrib) module (95) (Fig. 2A). Cell polarity requires the segregation of each component to its appropriate apical or basal compartment (96, 97). Polarity proteins, adhesion molecules, and mechanochemical pathways closely interact to maintain tissue organization and homeostasis (98). Signaling from the BM through adhesion molecules appears to be essential for epithelial polarization in many developmental and three-dimensional (3D), cell-based model systems (99, 100).

In contrast to simple epithelia, the stratified epithelium of the skin displays apico-basal polarization, with differential expression of polarity proteins across the epidermal layers (62, 101, 102). In the epidermis, tight junctions are not present in the basal cells, but form only in the differentiated granular layer, where they establish a continuous barrier to molecules and their paracellular passage from the exterior (Fig. 2) (103). Epithelial monolayers polarize with apical enrichment of junctional complexes, while stratified epidermis exhibits polarity that is primarily created along the basal-to-apical axis of the tissue (101). Perhaps this can be explained by the fact that cell polarity in basal keratinocytes carries the obvious advantage of being necessary for spindle orientation and asymmetric cell divisions, which result in the transition from a proliferative basal cell to a suprabasal cell committed to differentiation (104). Among the pathways involved in this event, such as Notch and EGFR signaling, integrin-based contacts with the BM appear to contribute to the basal keratinocyte polarity and proliferation/differentiation decision (73, 105). In addition, integrin-mediated interactions of cells with the BM—lead to its further deposition, supramolecular organization, and remodeling—factors that sustain polarity (106-109).

Studies using cultured cells have long suggested that laminins may function as polarity cues for developing epithelia (110). Early studies highlighted the involvement of a β1 integrin-dependent PI-3K signaling and Rho GTPase Rac1-mediated LM assembly mechanism at the Madin–Darby canine kidney (MDCK) apical cell surface (106, 111-113), revealing that epithelial cells coordinate polarity with tissue architecture. Impairment of LM assembly by Rac1 inactivation, or by inhibition of β1 integrin function, leads to an inversion of apical–basal polarity (106, 112). Conditional ablation of the GTPase Cdc42 in the mouse epidermis does not dramatically affect polarity, but it results in later defects in the BM (114). Members of the Par system are also involved in the β1 integrin-dependent LM assembly and remodeling mechanisms (99). Genetic studies in *Caenorhabditis elegans* reveal that LM provides a crucial cue that determines the apical localization of the Par3 complex at the opposite apical surface of epithelia during pharyngeal cyst development (115). Par3 is also apically located in basal keratinocytes, and it appears to control epidermal integrity through Rho-dependent keratinocyte mechanics (116, 117). Thus, β1 integrin is clearly required for cell polarity maintenance in the epidermis (73). In mouse epidermis, β1 integrin loss-of function experiments have revealed defective peripheral keratinocyte membrane protein polarization, and randomization of cell division orientation (72, 73, 116). Thus, β1 integrin is clearly required for cell polarity maintenance in the epidermis. Additionally, the loss of any of several β1 integrin-mediated focal adhesion proteins breaks the cytoskeleton–BM linkage, disrupts the BM, and impairs intracellular apical–basal orientation (72, 73, 118). Experimental ablation of binding partner integrin-linked kinase (ILK) (119), ILK-binding protein PINCH1 (120), integrin binding kindlin (121), and actin-bundling protein T-plastin (Pls3) (122) have all resulted in defects of BM assembly and epidermal homeostasis.

The β1-integrin-binding protein kindlin-1, which is highly expressed in the epidermis (123), is reportedly essential for basal keratinocyte adhesion, polarity in the epidermis, and epidermal BM organization (62, 124). Mutations in the kindlin-1-encoding FERMT1 gene cause Kindler syndrome, an autosomal recessive disease characterized by blistering at the DEJ, followed by skin atrophy (125). Patients with Kindler syndrome exhibit a thick and disorganized BM (126). Genetic ablation of

kindlin-1 in mice leads to reduced epidermal proliferation and thickness (127), and kindlin-1-deficient skin cells display a loss of polarity and proliferation (128). Kindlin appears to contribute to integrin clustering and trafficking and adhesion turnover (129, 130). Establishment and maintenance of cell polarity also rely on the transport of newly synthesized and recycled proteins to their correct destinations (131, 132). Studies of the *Drosophila* follicular epithelium have revealed that the phospholipid phosphatidylinositol 4,5-bisphosphate (PIP2) plays specific roles in controlling the polarized secretion of BM components, such as collagen IV and perlecan (133). β4-integrins have also been implicated in the regulation of LM-triggered polarity signals in epithelial cells, particularly in 3D *in vitro* culture systems (107, 134, 135). In mammary epithelial acini, β4 integrin-induced formation of hemidesmosomes promotes cell polarity and resistance to apoptosis (134). In human keratinocytes, integrin α6β4 interacts with its intracellular partner, BPAG1e, promoting cell polarity through modulation of Rac1 activity (135). Likewise, in a model in which physiologically relevant electric fields are applied to human keratinocytes, the binding of α6β4 integrin to LM-332 supported lamellipodium-associated Rac1-dependent persistent directional migration and polarity (136). The cellular receptor dystroglycan, which localizes to hemidesmosomes to strengthen dermal–epidermal anchorage, might regulate cell polarity through its interaction with Par-1 (137, 138). This has also been observed in the *Drosophila* follicular epithelium, where the interaction between dystroglycan and perlecan is required to maintain cell polarity (139). Together, these data show that signals coming from the BM contribute to the maintenance of epithelial polarity in the epidermis, and contribute to its renewal.

## DYNAMICS OF THE DERMAL-EPIDERMAL JUNCTION DURING SKIN WOUND RE-EPITHELIALIZATION

Numerous studies conducted in mammalian animal models, in humans, and in *in vitro* and *ex vivo* human tissue models have revealed that during the skin repair process, regeneration of a fully functional epidermis depends on accurate reconstitution of the DEJ, and on the subsequent differentiation of keratinocytes (140). Data from a three-dimensional (3D) organotypic full-thickness *in vitro* skin wound model by two-step time-lag fluorescence staining and data collected from an epidermal scratch model in the dorsal skin of a fluorescent mouse each showed that in minor epidermal injuries, wound edge keratinocytes collectively migrate to re-epithelialize the wound as the first step in re-establishing tissue integrity and barrier function (141, 142). Two studies conducted in a mouse wound model revealed that the migrating and non-proliferative epidermal leading edge is followed and supplied by a highly proliferative zone (143, 144). In this context, cell mechanics and the actomyosin cytoskeleton are fundamental for allowing epithelial closure (145). Molecular data obtained from studies on invertebrates coupled with dynamic studies on vertebrates including mammalian models have resulted in understanding that cell proliferation, division, orientation, and polarized migration must be coordinated within a collective movement that promotes the directional expansion of the epithelium (98). These findings, combined with additional state-of-the-art *in vitro* models, have concluded that the integrin-mediated interactions with the substrate, as well as the substrate's rigidity, control the speed and efficiency of these events (146, 147).

Lessons learned from mammalian models and work in humans have provided insight into the dermal repair and epidermal regeneration processes of wounds in physiological conditions (148). The early steps of wound closure are stimulated by cytokines and growth factors produced by both inflammatory and skin cells (149). The migrating keratinocytes first encounter a provisional matrix, mainly comprising fibrin, plasma fibronectin, vitronectin, and platelets (150). The keratinocytes then deposit newly produced ECM proteins along their migration path, including LM-332, cellular fibronectin, tenascin C, and matrix metalloproteinases (MMPs) (140, 143). Fibronectin is ubiquitously expressed in human tissue and it has long been known to be a major component in wound healing (151). Alternative spliced variants of cellular fibronectin, deposited by migrating keratinocytes or present in the granulation tissue, contribute to the regulation of multiple biological functions that benefit the healing process—notably including cell migration, inflammatory response engagement, and matrix assembly properties (140, 150, 152).

The matricellular protein tenascin C is undetectable in most healthy adult tissues but is transiently re-expressed during wound healing and dynamic tissue remodeling (153). This large hexameric extracellular glycoprotein with cell signaling properties is rapidly induced at sites of inflammation both underneath the regenerating epidermis as well as in the stroma (154, 155). *In situ* hybridization revealed that keratinocytes are a source of tenascin-C (156), however no function in skin wound re-epithelialization was demonstrated in a tenascin C knockout mice mouse model (157). In contrast, analysis of the repair process analysis of incision-injured corneas in tenascin C KO mice, revealed that tenascin C was required for the primary healing of the corneal stroma by modulating wound healing-related fibrogenic gene expression in ocular fibroblasts (158). Tenascin C is known to accumulate in chronic pathological conditions (159) and a recent study in skin from patients with systemic sclerosis and in mouse models of organ fibrosis revealed its involvement in the TOLL-like receptor 4-dependent fibroblast activation (160).

LMs play an active role in the epidermal regeneration process in addition to their structural and biological activity in the DEJ in both human tissue and mouse wound models. As revealed in mice, in human tissues, and *in vitro* tissue-engineered human skin models, increased LM-332 expression is one of the earliest events in wound re-epithelialization, and LM-332 is among the first BM components laid down onto the *in vivo* wound bed by migrating keratinocytes (140). Many studies over the years with human cells, in *vitro* 3D models, as well as *in vivo* mouse wound models have shown that keratinocytes of the regenerating tongue remain in contact with the newly deposited LM-332, and that both integrins α3β1 and α6β4 participate in this interaction and in the migration process (55, 56, 161-167). Furthermore, recent studies using mass spectrometry-based proteomics on conditioned medium from cultured human cells suggested that the α3β1 integrin is a regulator of the secretome in keratinocytes (168), and the α3β1 integrin/LM-332 interaction reportedly triggers gap junction assembly and promotes intercellular communication in human wounding epidermis both *in vitro* and *in vivo* (169). Using high resolution imaging of the immortalized human keratinocyte HaCaT cell line, recent data showed a coordinated dialogue between actin stress fiber-associated focal adhesions and keratin-associated hemidesmosomal structures (59), opening up perspectives for clarification of the complex cross-talk mechanisms between focal adhesions and hemidesmosomes that appears to take place during keratinocyte migration (170-173).

The complexity of these mechanisms is increased by the fact that other proteins of the provisional matrix, may possibly have an influence on matrix organisation and signals to the cells. A study of pulmonary epithelial regeneration highlights this aspect by demonstrating the important impact of fibronectin on LM-332-mediated cell migration signals (174). Importantly, the mesh-like network of keratinocyte-derived LM-332 deposits promote different biological activities from those observed with purified, unassembled molecules (175).

Little is known about the potential existence of cooperative mechanisms between LM-332 and tenascin C. An interaction between these proteins in corneal repair revealed that these proteins were secreted and co-deposited in early adhesion plaques of *in vitro* wounded corneal epithelial cells (176). The complexity is further amplified by the fact that the molecular form of LM-332 in the provisional matrix differs from that in the mature DEJ, in that LM-332 is initially present in its precursor form (38, 177-179). The precursor form could exhibit additional, transient capacity for molecular interaction with other matrix components or cellular receptors, as has been proposed for syndecans, that promote the formation of actin-based, protrusive adhesion structures and activation of MMP expression (180-182). Full-length LM-332 was described as a ligand for syndecan-1 (180, 181), a cell surface receptor that is redistributed from the lateral to basal surface of keratinocytes in the migrating epidermal tongue (183). The transmembrane glycoprotein CD44, which is also localized at the basal surface of migrating keratinocytes (183), was recently identified as a molecular partner of precursor LM-332 in primary keratinocytes (25, 184). In addition to its functions in cell adhesion and signaling (185), CD44 forms a scaffold at the cell surface for the assembly and activation of various MMPs that ultimately modulate cell migration. CD44/LM-332 interaction in migrating keratinocytes could trigger a cascade for the recruitment and activation of MMPs that are essential for the regenerative processes (184). This hypothesis is supported by a study showing that transformed keratinocytes lacking the syndecan-1/CD44-binding domain of LM-332 exhibit deficient MMP activity and show decreased invasive capacity of carcinoma cells in nude mice (186). Collagen VII is another component found in the early provisional epidermal matrix that plays a critical role in skin wound closure (187). Wound healing studies in two recessive dystrophic epidermolysis bullosa (RDEB) mouse models, have demonstrated the instrumental role of collagen VII in both wound re-epithelialization and maturation of granulation tissue (187). The lack of collagen VII in keratinocytes delayed re-epithelialization through alteration of LM-332 deposition and the modulation of integrin α6β4 signaling.

Monitoring partial-thickness wound closure in human volunteers revealed that once the epidermis has covered the wound bed, the proteins of the DEJ appear to be sequentially enrolled from the margins to the center of the wound (188). It has been proposed that mature hemidesmosome formation may coincide with LM-332 maturation, an event that could occurs in later phases of wound healing (38, 54). The examination of wound re-epithelialization in knock-in mice expressing a functional, non-sheddable collagen XVII mutant highlighted the role of collagen XVII ectodomain shedding as a dynamic modulator of proliferation and motility of activated keratinocytes through tight coordination of α6β4 integrin-LM-332 interactions (189). In contrast, LM-511 is not deposited by migrating keratinocytes (11, 190) and seems to accumulate in the matrix of *in vitro* cultured confluent keratinocytes (190). However, a role for LM-511 is suggested, since its expression was correlated with the ability of keratinocytes to regenerate an intact epidermis (191, 192). The observation of an inhibitory role for LM α5 in basal keratinocyte proliferation and motility in a keratinocyte-specific LM-α5 knock out mice would coincide with its late expression during the re-epithelialization process (90).

Collagen IV is also detected in late phases of epithelialization, most likely anchoring the LM-511 network and contributing to BM assembly and stabilization (193, 194). Both nidogens and perlecan intervene in the BM maturation phase, probably through their bridging and stabilizing functions (183, 190); however, it has also been proposed that nidogen-1 has a potential unidentified biological function toward wound resolution (195). As the major proteoglycan of BMs, perlecan acts as a reservoir for growth factors, allowing their diffusion, and playing a role in keratinocyte survival and terminal differentiation (196, 197). In coordination with LM332, perlecan deposits contribute to reinforcing keratinocyte anchorage (190).

**EPIDERMAL DEFECTS LINKED TO BASEMENT MEMBRANE ABNORMALITIES**

While the dynamic nature of the BM is essential for tissue development, homeostasis, and repair, BM dysregulation is associated with various human diseases that affect many organs, including the skin (11, 22, 29, 198, 199). The importance of the BM in dermal–epidermal cohesion is illustrated in both inherited and acquired blistering diseases. Congenital inherited epidermolysis bullosa (EB) describes a group of devastating disorders characterized by mutations in any of the genes encoding anchoring complex components, leading to dermal–epidermal breakages involving various morphological levels of separation within the DEJ (200-202). Although all EB types and subtypes exhibit trauma-induced skin blistering and fragility, each type presents its own spectrum of clinical manifestations in terms of the extent of skin lesions and organ dysfunction. Anchoring complex proteins are also the targets of autoimmune diseases called pemphigoid diseases (203, 204). Characterization of the auto-antibodies responsible for epidermal dissociation in these pathologies has contributed to the elucidation of some BM protein interaction mechanisms (203, 205, 206). Stevens–Johnson syndrome and toxic epidermal necrolysis are other life-threatening diseases that involve breakdown of the BM in skin and mucous membranes (207). Both conditions are acute and rare, characterized by widespread epidermal necrolysis and sloughing, and classified by the extent of the detached skin surface area (208). Although a variety of etiologies, including infections and underlying malignancies, have been implicated as potential causes, drugs are the major causative agent leading to cell-mediated cytotoxicity and epidermal apoptosis. Apoptosis-inducing stimuli include specific CD8$^+$ cytotoxic lymphocytes, the Fas-Fas ligand pathway of apoptosis, and the granule-mediated exocytosis and tumor necrosis factor-a pathway (209). This apoptosis results in hemidesmosome anchorage failure, and the accumulation of serous fluid over time, leading to blister formation and massive de-epithelialization, inflammation, and necrosis.

The BM is a dynamic structure that can change through either protein synthesis/degradation, non-enzymatic post-translational modifications, or reorganization of its existing components (210). Such changes often lead to alterations in other BM features, including thickness and mechanical properties. Such modifications can be observed during aging, leading to consequences at the level of the epidermis—an aspect that was recently described in a detailed review (7). The involvement of DEJ components as a niche for epidermal stem cells is of paramount importance during aging. For example, examination of skin of subjects at various ages showed that intrinsic, age-related defects in perlecan and UVB-induced damage of LM-511 negatively impact expression of melanoma chondroitin sulphate proteoglycan (MCSP)-positive basal epidermal stem cells (78) and keratin-15-positive cells (78, 211), leading to a reduced basal cell density in the epidermis. Studies in mice highlighted that age-related reduction of collagen XVII expression causes hemidesmosome alterations, keratinocyte detachment from the DEJ, and hair loss (63, 212). Moreover, the age-related loss of LM-332 expression in the stem cell niche may cause impairment of the epidermal differentiation process (66).

Aging is characterized by the overexpression and activation of MMPs that target and degrade DEJ components, disturbing their mechanical and biological properties (213). In a 3D skin equivalent model, MMP inhibitors are able to restore the UVB-induced defective DEJ leading to restoration of epidermal homeostasis (214, 215). Age-related changes in BM composition may progressively modify its mechanical properties, causing an impaired epidermal response to mechanical forces (216, 217) and deleterious consequences on the fate of progenitor cells (7, 218, 219). Notably, reduced expression of collagen XVIII in the DEJ of an *in vitro* skin equivalent model has been shown to result in decreased elasticity and strength of the tissue (220). A recent study has demonstrated that stiffening of the aged hair follicle stem cell niche impacts chromatin accessibility, as well as the epigenetic state of key activation, self-renewal and differentiation genes (221), thereby attenuating stem cell activation and tissue regeneration.

BM component anomalies at any stage of the wound healing process can lead to failed or delayed resurfacing and wound closure compromise (222). The clearest indicator of a chronic wound is the failure to re-epithelialize, as dysfunctional keratinocytes continuously proliferate but fail to have an adequate granulation tissue substrate for migration (223). It is clearly established that inflammation induces excessive protease activity that promotes degradation of the adhesion proteins required for normal wound healing (224, 225). For example, degraded fibronectin and tenascin-C are found in the wound fluids of chronic ulcers in humans (224, 226, 227). In venous ulcers, depletion of stem cell niche signaling leads to stem cell depletion, migration defects, and abnormal differentiation (226, 228, 229). Alterations of integrin and BM components can also prevent cell migration (230). Moreover, the deregulation of LM-332 expression and maturation leads to excessive release of the C-terminal globular domains G4 and G5 of its α3 chain (180, 231).

Little is known about BM reconstruction in late phases of wound healing, and whether cells can rebuild a BM identical to the original in all respects. Studies of human cornea regeneration have revealed a slow recovery of BM cell adhesion properties (232). *In vivo* imaging of epithelial wound healing in the cnidarian *Clytia hemisphaerica* has demonstrated that the interaction between epithelial cells and the BM establishes the closing mechanism, and that lamellipodia-dependent cell crawling requires an intact BM (233). Analysis of BM repair in the larval epidermis of the fruit fly, *Drosophila melanogaster*, has revealed that BM components are repaired, leaving a visible scar due to excessive deposition, and that the assembly mechanisms involved in repair may differ from those involved in development (234). Although the parallelism between invertebrate and vertebrate healing mechanisms is difficult, these data indicate that BM reassembly and remodeling play an essential role in the regeneration/repair process.

During aging, the persistent inflammatory context, and the emergence of detrimental, post-translational modifications of long-lived ECM proteins, lead to deficiencies in molecular network assembly, giving way to fragile scars prone to wound recurrence and reopening (29). Moreover, aging BMs show increased stiffness due to the structural and biomechanical changes caused by advanced glycation end product-induced crosslinks in the collagen IV and LM networks (235, 236).

BM anomalies can be associated with exacerbated stimulation, as is the case in excessive wound healing that causes impaired epithelialization and results in hypertrophic scars or keloids (237). In a study conducted using biopsies of patients with hypertrophic scars, the findings indicate that the defective expressions of collagen IV, LM-332, and β4 integrin in the BM reduce the attachment of basal progenitor cells, and induce them to adopt a proliferative phenotype (238). Keloids have been shown to display epidermal changes linked to their hyperproliferative state (239, 240), as well as ultrastructural alterations of the BM, displaying fewer hemidesmosomes (237).

Finally, the BM also undergoes pathological remodeling or mechanical stresses in the settings of cancer and tumor invasion—pathological situations that are widely described in recent reviews (20, 69, 241).

**CONCLUSION—FUTURE PROSPECTS**

As an exposed tissue that is easy to manipulate and observe, the epidermis was one of the first targets for cell isolation, *in vitro* tissue engineering, and *in vivo* experimental gene transfer (242). A number of skin diseases, particularly inherited skin disorders, have benefitted from advances in molecular genetics and biotechnology, and subsequently in gene therapy approaches (243). Successful and extremely promising results have been achieved with autologous transgenic keratinocyte cultures, regenerating an entire, fully functional epidermis in a 7-year-old child suffering from a devastating, life-threatening form of junctional EB (244). Gene therapy for wound healing is currently designed to boost factors that are known to assist with the wound-healing process (140).

Various strategies have been used to mimic BM complexity when bioengineering organotypic cultures (7, 245-248). Decellularized xenografts show promise for wound healing management—retaining the native ECM structure, comprised of dermal collagen and the structural, adhesion, and signaling molecules of the BM (249). However, more investigation is needed before synthetic BMs can be produced. To evaluate the BM's barrier function and mechanical properties, investigators have fabricated natural BMs in 3D spheroids of human mammary epithelial (MCF-10A) and breast cancer cells (MDA-MB-231) (250, 251). Skin organoid/spheroid technology will help elucidate the cellular mechanisms of epidermal homeostasis and serve as a platform for studying the basic biology of skin diseases and their underlying genetic alterations (82, 252). Epidermal organoids that resemble the interfollicular epidermis were recently established from adult mouse skin, including proliferative cells in the basal layer and differentiation toward the cornified envelope (253). Understanding the mechanisms underlying BM formation and reconstruction is important for optimizing and improving tissue engineering strategies.

Our review focused on the chemical cues provided by BM proteins to cells, with BM macromolecules acting as ligands for cell surface receptors. ECM mechanical properties, represented by nanotopography and elastic modulus, can modify cell behavior independently of chemical signals (254, 255). This is increasingly important in tissue engineering, where a biomaterial scaffold must rapidly induce cellular intake and tissue regeneration signals. We must understand the mechanical impact of the provisional BM and its remodeling on the successive steps of epidermal regeneration (256, 257). Much recent knowledge about BMs has come from studies on model organisms. Studies in *Drosophila* have demonstrated the ECM's evolving structure and mechanical patterning, allowing BM stiffness to regulate epithelial morphogenesis, guide cells, and shape organs (258, 259). Increasing knowledge of epidermal homeostasis and regeneration was obtained with the development of substrates containing nano- to micron-scale features to mimic those found in the native skin BM (7). Improving and optimizing the topographic cues that affect keratinocyte morphology, migration, proliferation, differentiation, and protein expression is a promising avenue for the development of innovative materials for skin regeneration.

**ACKNOWLEDGMENTS**

We thank Jeffrey M. Davidson from Vanderbilt University Medical Center for his critical reading of the manuscript. The authors' research was funded by a Pack-Ambition Rhône Alpes Region 2020 grant to P.R. (3D-Healing).

**DISCLOSURES**
The authors declare that they have no conflicts of interest with the contents of this article.

**AUTHOR CONTRIBUTIONS**
P.R. wrote and revised the manuscript. C.L. contributed to the wound re-epithelialization paragraph. G.R. and P.R. prepared figures. C.L., G.R., and P.R. approved the final version of the manuscript.

FIGURE LEGEND

FIGURE 1

**Figure 1.** Molecular anchorage of the epidermis and basement membrane framework. (*A*) General view of the molecular networks at the interface between the epidermis and the dermis. The epidermis, an epithelium mainly comprising keratinocytes, is connected to the underlying dermis via a structure called the basement membrane (BM). A dynamic network of laminin (LM)-511 molecules, and a mechanically resistant collagen IV scaffold, are the two foundational skeletons of the BM, linked together by stabilizing and regulatory proteins, such as nidogens and perlecan. The LM-511 network binds to basal keratinocytes through α3β1 integrin-mediated focal adhesions, which are linked to the intracellular actin cytoskeleton

through interacting intermediate proteins, such as talin, kindlin, and focal adhesion kinase (FAK). To this basic framework, are added molecular straps that firmly anchor the epidermis. These straps comprise extracellular matrix macromolecules, such as LM-332 and LM-311, and transmembrane proteins such as collagen XVII and bullous pemphigoid antigen (BP) 180 and integrin α6β4, all localized at the basal surface of keratinocytes. Their clustering forms hemidesmosomes: adhesion structures linked to the intracellular keratin cytoskeleton through bridging elements, such as plectin and BP230. On the basal cell surface, this complex is connected to the dermis by collagen VII fibrils, which attach to LM-332 and intertwine in the collagen I fibers of the superficial dermis. A pattern created from collagen XVIII reinforces these boundaries. Fibroblasts are the main cell type in the dermis responsible for the production of ECM. (*B*) Enlarged view of the keratinocyte-BM interface. Right: focal adhesions mediated by the interaction between integrin α3β1 and LM-511, and linked to the actin cytoskeleton. Center: LM-511 in contact with transmembrane dystroglycan. Left: the numerous extra- and intracellular molecular interactions leading to hemidesmosome formation—with integrin α6β4 at the center.

**Figure 2.** Schematic representation of the barriers in simple epithelia and stratified epidermis. (*A*) Simple epithelia comprise one layer of cells that attach to the basement membrane by focal adhesions and hemidesmosomes, and to adjacent cells via tight junctions, adherens junctions, desmosomes, and gap junctions as indicated. Tight junctions contribute to the maintenance of apical-basolateral polarity. The plane of the mitotic spindles aligns perpendicularly to the basement membrane, allowing lateral expansion of the cells. The tight junctions and adherens junctions separate apical and basolateral domains, and they promote configuration of the apico-basal polarity. The Crumbs complex—comprising the proteins Crumbs, PATJ, and Pals1—localizes to the apical membrane. The Par complex, which includes aPKC, Par3, Par6, is laterally localized, where it interacts with TJ proteins. Scribble, Lgl, and Dlg form the Scribble complex, which is basolateral. The localization and mutual interactions between polarity complexes and cell/cell junction proteins are important for the establishment and maintenance of apico-basal polarity. Mutual interactions, most notably phosphorylation, regulate the segregation of the proteins to the apical or basolateral side. (*B*) The mammalian epidermis is a stratified squamous epithelia in which the intercellular junction proteins are polarized across multiple layers of the epidermis. Adherens junctions, desmosomes, and Gap junctions attach cells to each other, while integrins present in focal adhesions and hemidesmosomes attach basal keratinocytes to the basement membrane. Tight junctions are found in the upper spinous layers. Dividing cells of the basal layer have either the spindle plane parallel to the basement membrane to allow asymmetric division of basal cells and vertical extension of the basal layer or perpendicular to the basement membrane to allow symmetric division of stem cells.

Figure 1

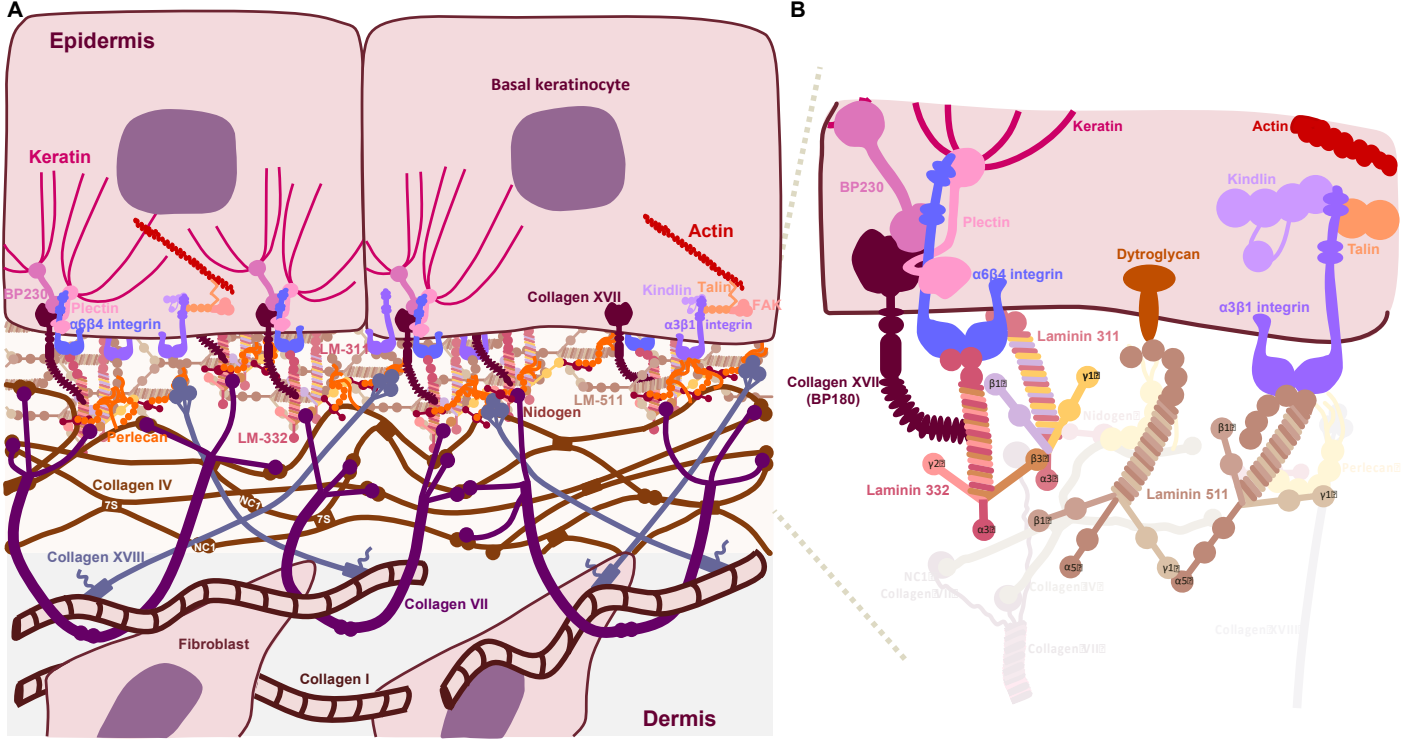

Figure 2

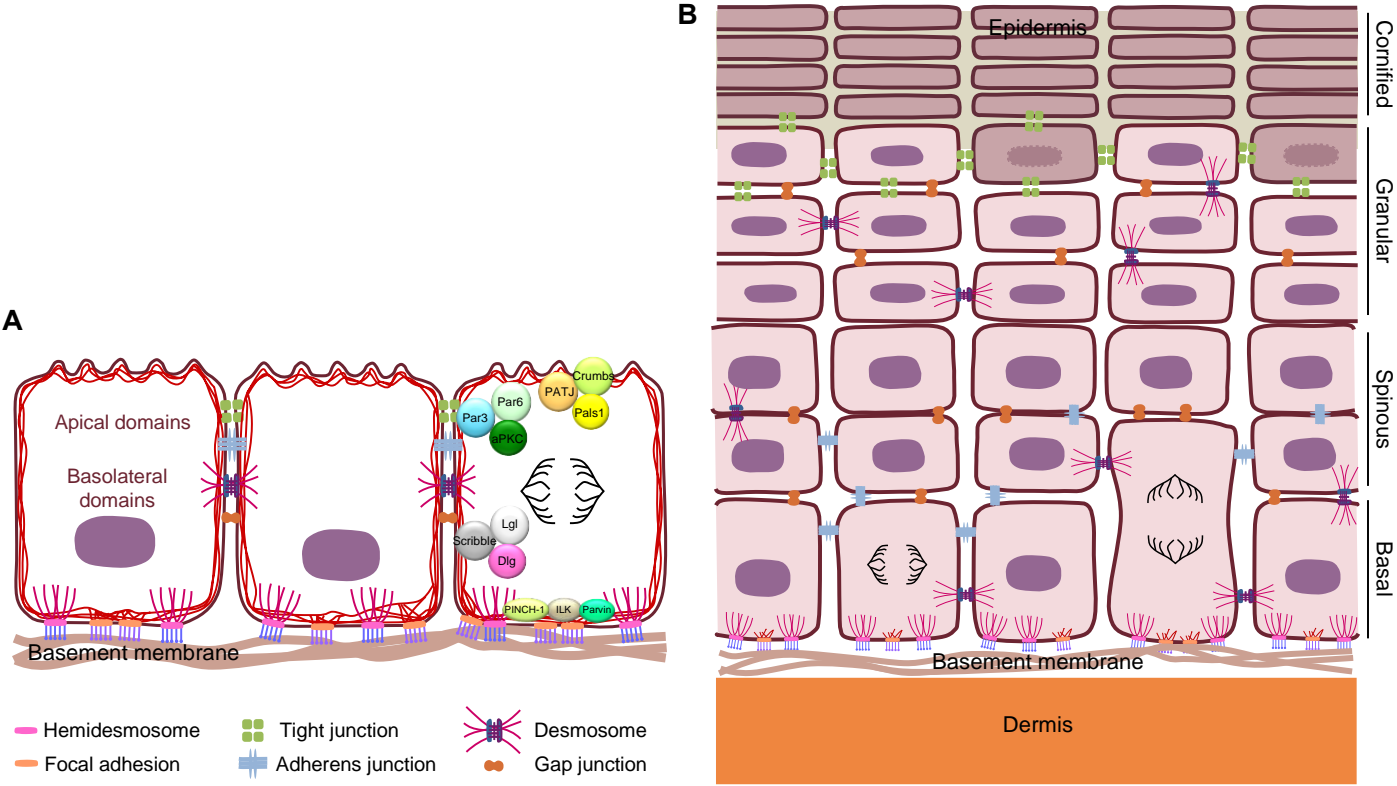